\documentstyle[12pt]{article}

\newlength{\dinwidth}                                                           
\newlength{\dinmargin}                                                          
\newcommand{\bfg}{\begin{figure*}}                                              
\newcommand{\efg}{\end{figure*}}

\begin{document}                                                                

\begin{center}
{\it Talk given at the DESY Workshop "PHYSICS at HERA", Hamburg, Germany,\\  
     29-30 October, 1991}
\end{center}      
                                                                        
\vspace{1cm}                                                                    
\begin{center}                                                                  
{\LARGE \bf TERAD91:}    \\[1em]                                                 
                                                                                
{\Large \bf A Program package for the calculation of}\\[1em]                    
                                                                                
{\Large \bf the cross sections of deep inelastic NC and CC}\\[1em]              
                                                                                
{\Large \bf scattering at HERA}\\[1em]                                          
                                                                                
                                                                                
  \vspace{8mm}                                                                  
                                                                                
  \begin{large}                                                                 
Arif A.\ Akhundov${}^1$, Dmitri Yu.\ Bardin${}^2$, 
\\                                           
L.\ Kalinovskaya${}^3$ and T.\ Riemann${}^4$                                   
  \end{large}                                                                   
                                                                                
  \vspace{5mm}                                                                  
                                                                                
   ${}^1$ Institute of Physics, Azerbaijan Academy of Sciences,\\  
         pr.~Azizbekova 33, 370143 Baku, Azerbaijan \\           
   ${}^2$ Joint Institute for Nuclear Research, Joliot-Curie 6, 
141980 Dubna, Russia\\                        
                                    
   ${}^3$ Gomel Polytechnical Institute, October ave.~48,  
          246746 Gomel, Belarus \\                                             
   ${}^4$ DESY - Institut f\"ur Hochenergiephysik,
         Platanenallee 6, D-15738 Zeuthen, Germany \\                 
                                                                                                                                                            
\vspace{5mm}                                                                    
\end{center}                                                                    
\begin{quotation}                                                               
\noindent                                                                       
{\bf Abstract:}                                                                 
We present a brief description of the recently developed Fortran code           
{\bf TERAD91} for semi-analytical calculations of the double differential         
cross sections of $\rm NC$ and $\rm CC$ deep inelastic electron-proton          
scattering and of some related observables. The code is                         
mainly intended for the calculations at HERA or other $ep$-colliders            
but may be also used for similar processes like muon-proton scattering          
in fixed-target experiments.                                                    
\end{quotation}                                                                 
                                                                                
\vspace{10mm}                                                                   
%
                                                                                
\section{Program Summary}                                                       
                                                                                
\begin{list}{}{%
\setlength{\leftmargin}{68mm}%
\setlength{\labelwidth}{55mm}}                                                  
\item[Title of the program:\hfill]                                              
                           TERAD91                                              
\item[Version:\hfill]                                                           
                           version 2.10, 3 Oct.\ 1991                           
\item[Computer:\hfill]                                                          
                           IBM-3090, DESY                                       
\item[Program language:\hfill]                                                  
                           FORTRAN-77                                           
\item[Number of program lines:\hfill]                                           
                           about 11,000                                         
\item[Other programs used:\hfill]                                               
                           DIVDIF, interpolation routine from the CERN          
                           library KERNLIB.                                     
\item[External files needed:\hfill]                                             
                           None.                                                
\item[Method of solution:\hfill]                                                
                           Numerical integration.                               
\item[Typical running time:\hfill]                                              
                           Depending on the actual requirements                 
                           like parameters, requested contributions,            
                           see text.                                            
\end{list}                                                                      
                                                                                
\section{Introduction}                                                          
                                                                                
{\bf TERAD91} is a software product by the Dubna-Zeuthen                        
Radiative Corrections Group (DZRCG).                                            
It is intended for calculations of Deep                                         
Inelastic Scattering (DIS) cross sections and related observables               
(cross section ratios, asymmetries etc.) at HERA.                               
Its typical output                                                              
is the double differential cross section of DIS                                 
with account of QED and electroweak (EW) radiative corrections                  
as chosen by flags:                                                             
\begin{eqnarray}                                                                
\frac{d^2 \sigma}{dxdy} ,                                                       
\end{eqnarray}                                                                  
(in nanobarn) with $x$ and $y$ being some scaling variables                     
- either leptonic or mixed or hadronic.                                         
For a brief description of the basic theoretical issues and definitions         
see \cite{contvol1}.                                                            
                                                                                
{\bf TERAD91} was created during the 1990-91 Workshop on Physics at HERA.       
It accumulates about 15 years experience in the field                           
of DIS collected by the DZRCG. Formally,                                        
it originates from four different codes:                                        
{\bf TERAD90} \cite{teradtex}, {\bf DISEPNC} \cite{disepnc},                    
{\bf DISEPCC} \cite{disepcc} and {\bf DIZET} \cite{dizet}.                      
                                                                                
{\bf TERAD90} was its prototype, created just before the Workshop.              
Its roots date back till the mid seventies \cite{oldpreps,oldpub}.              
In the eighties, the TERAD approach was used                                    
for the analysis of muon DIS data by the BCDMS experiment                       
\cite{terad86,bcdms}.                                                           
                                                                                
{\bf DISEPNC and DISEPCC} are codes developed in the late eighties              
for the QPM calculations of DIS at HERA \cite{disepnc,disepcc}.                 
A similar approach was used earlier for the analysis                            
of neutrino DIS data by the CDHSW \cite{cdhsw}                                  
and CHARM-I \cite{charmI} experiments.                                          
                                                                                
{\bf DIZET} is our electroweak library, used before for LEP-I                   
physics  \cite{dizet}.                                                          
It allows to calculate $\Delta r$, and the complete                             
$ \cal O (\alpha)$ corrections,                                                 
with inclusion of some leading higher order terms                               
connected with the top-quark, to $Z,W$ widths,                                  
and to the weak form factors for cross sections                                 
in the $s$- and $t$-channels.                                                   
These form factors depend on the kinematics ($s,t,u$-invariants),               
and on the fermion type of the scattering particles.                            
                                                                                
Some theoretical basis of the TERAD approach is presented in the                
contributions of the working group on radiative corrections in these            
proceedings                                                                     
\cite{contvol1}. Here we concentrate on the description                         
of the code itself assuming that the reader is aware of the relevant            
physical problems and of the terminology.                                       
We present a description of its                                                 
{\tt MAIN} routine and of                                                       
{\tt SUBROUTINE SET} where all flags are collected                              
which should be set by a user.   \\                                             
                                                                                
\section{MAIN program of TERAD91}                                               
                                                                                
The {\tt MAIN} routine of {\bf TERAD91} consists of a                           
{\tt CALL} of {\tt SUBROUTINE SET}                                              
which sets various user flags followed by a                                     
{\tt CALL} activating one of the calculational {\bf chains},                    
which computes some observable (e.g. cross section)                             
or some other physical quantity (e.g. an electroweak form factor).              
Normally only one chain is active,                                              
{\tt CALL}s to others are commented.                                            
We recommend to work only with one chain at a time.                             
Otherwise the running would be too cumbersome as                                
some of the calculations are time comsuming and                                 
default outputs are lengthy. Moreover, a destructive cross influence            
of chains is not excluded.                                                      
Below we briefly describe each {\bf chain}. \\                                  
{\bf DXYLEP}             \\                                                     
{\tt SUBROUTINE DXYLEP}                                                         
calculates double differential cross sections in terms of usual                 
leptonic scaling variables $x,y$. For leptonic type variables,                  
{\bf TERAD91} is mostly elaborated.                                             
It calculates both $\rm NC$ and $\rm CC$ cross                                  
sections in both TERAD and DISEP approaches (we will refer to them as           
two different {\bf branches}).                                                  
For this reason the output of                                                   
{\tt DXYLEP}                                                                    
contains a lot of cross sections (Born, corrected by EW and/or QED)             
and corrections calculated by both branches.                                    
The resulting cross section and correction for $\rm NC$ reaction is a           
mixture of numbers, calculated by both TERAD and DISEP branches.                
Because of two-fold numerical integrations, involved                            
in the calculation of cross sections in this case,                              
the running with {\tt DXYLEP} is rather slow                                    
and takes about 70 minutes for a calculation in 96 kinematical points           
(default option) on the DESY IBM.        \\                                     
{\bf AXYLEP}                   \\                                               
{\tt SUBROUTINE AXYLEP} calculates a                                            
double differential polarization asymmetry in leptonic variables $x,y$.         
All calculations again are performed by both branches; since now                
one has to calculate each cross section twice, for two polarizations,           
the calculational time needed is approximately doubled.                         
When using this chain, one should have in mind that                             
initialization of polarizations is done inside {\tt AXYLEP}.                    
Also electroweak                                                                
parameters could be re-initialized there in order to investigate                
the sensitivity                                                                 
of the asymmetry to the EW parameters.  \\                                      
{\bf DXYMIX}                      \\                                            
{\tt SUBROUTINE DXYMIX} calculates                                              
double differential cross sections in terms of mixed scaling variables          
$x_m,y_m$.  \\                                                                  
{\bf DXYHAD}                      \\                                            
{\tt SUBROUTINE DXYHAD} does the same in terms of hadronic variables            
$x_h,y_h$ (for their definitions see \cite{contvol1,teradtex}).                 
                                                                                
For mixed and hadronic variables, the {\bf TERAD91} package contains            
only the TERAD approach for only the $\rm NC$ reaction.                         
For this reason the outputs of {\tt DXYMIX} and {\tt DXYHAD} are                
very short: only the $\rm NC$ Born                                              
cross section and total correction are printed for                              
each chosen $(x,y)$ pair. As only one numerical integration is left for         
the calculation of the cross section, these chains are much faster as           
compared to the first two. Each of them require less than 1 minute for          
the calculation in 96 default kinematical points.                               
                                                                                
The three subroutines {\tt DXYLEP}, {\tt DXYMIX} and {\tt DXYHAD}               
are the main interfaces to the                                                  
core of the {\bf TERAD91} code - which realizes some                            
formulae describing differential DIS cross sections with account of             
QED and EW Radiative Corrections (EWRC) in some scaling variables $x,y$         
, see Section 3.2.1 of \cite{contvol1} and \cite{teradtex}.                     
But, {\bf TERAD91} can easily also be                                           
used to calculate many other observables, related to differential               
cross sections, and the discussed above                                         
{\tt AXYLEP} gives an example of a user interface for the                       
calculation of another observable, namely polarization asymmetry $A$.           
Similar to {\tt DXYLEP}, {\tt DXYMIX} and {\tt DXYHAD},                         
{\tt AXYLEP}                                                                    
also returns a double differential in $x$ and $y$ quantity.                     
                                                                                
The three subsequent interfaces exhibit                                         
examples of the calculation of integrated                                       
quantities: cross section ratio $R={\rm NC}/{\rm CC}$                           
({\tt RMIN1} and {\tt RMIN2})                                                   
and again the polarization asymmetry                                            
$A$ ({\tt AMIN1}) in a kinematical region                                       
restricted by some cuts.                                                        
All these subroutines deal                                                      
with definitions of cross sections in terms of leptonic variables and           
use the DISEP approach.                                                         
In {\tt RMIN1,2} and {\tt AMIN1}, corresponding observables are                 
constructed out of the Improved Born Approximated (IBA)                         
DIS cross sections, i.e. with EW but without QED radiative corrections.         
This is done to study their sensitivities to the variation of the EW            
parameters.                                                                     
Due to additional integrations over the restricted by cuts                      
kinematical region, these interfaces are rather slow. It is another             
reason why we use the DISEP approach there and restrict ourselves               
to the IBA. \\                                                                  
{\bf RMIN1} \\                                                                  
{\bf AMIN1} \\                                                                  
{\tt SUBROUTINEs RMIN1} and {\tt AMIN1} calculate $R$ and $A$                   
with cuts in $x,y$ and $Q^2$.  \\                                               
{\bf RMIN2}          \\                                                         
{\tt SUBROUTINE RMIN2}                                                          
calculates $R$ with a cut in $p_\perp^h$ and $Q_h^2$.    \\                     
All of them return $R$ and $A$                                                  
as functions of varying top-quark and Higgs masses.                             
Similar interfaces can be readily written by a user to calculate                
other integrated observables.                                                   
                                                                                
The three last interfaces are intended for                                      
a calculation of purely EW effects.                                             
They are indeed interfaces only to the EW library {\bf DIZET}.                  
They were used to produce some Tables and Figures for the                       
Workshop proceedings and are left in the                                        
{\bf TERAD91} as further examples of interfacing. \\                            
{\bf DIZETF} \\                                                                 
{\tt SUBROUTINE DIZETF}                                                         
returns a Table for the static EW parameters,                                   
$M_W$ and $ \sin^2 \theta_W$.\\                                                 
{\bf FORMFS} \\                                                                 
{\bf FORMFT} \\                                                                 
{\tt SUBROUTINEs FORMFS} and {\tt FORMFT} return some tables                    
illustrating the $s$- and $t$-dependences of EW form factors.                   
We will not describe them in detail here. \\                                    
                                                                                
\section{SUBROUTINE SET}                                                        
                                                                                
In this {\tt SUBROUTINE} we collected all flags used to initialize the          
{\bf chains}. The main interfaces                                               
{\tt DXYLEP}, {\tt DXYMIX} and {\tt DXYHAD}                                     
are initialized by {\tt SET} completely.                                        
Other interfaces may contain some additional initialization inside them;        
{\tt SET} prints also all flags.                                                
                                                                                
In {\bf TERAD91}                                                                
there are flags which act in both the TERAD and DISEP {\bf branches};           
but there are also flags inherent to only one of the branch.                    
The notion of default flags is not supported in {\bf TERAD91}.                  
Flag setting                                                                    
should be controlled by the user for each specific run. \\                      
                                                                                
\subsection{TERAD - flags}                                                      
                                                                                
{\tt MODEL}  \\                                                                 
{\tt MODEL=1}: quark distributions are used for the calculation of the          
structure functions,                                                            
(presently the Duke and Owens \cite{dukeowens} parametrization), \\             
{\tt MODEL=0}: in principle any parametrization can be used for                 
the eight structure functions                                                   
$F_1,F_2$, $G_1,G_2,G_3,H_1,H_2,H_3$. The user can                              
substitute instead of using {\tt SUBROUTINE SFF1} (for $F_1$) etc.              
any parametrization available to him/her.                                       
Now we use a merge of Brasse \cite{brasse},                                     
Stein \cite{stein} and  Duke-Owens \cite{dukeowens} fits (see                   
\cite{dimajacek} for more details). \\                                          
{\tt IVAR} \\                                                                   
{\tt IVAR=0}:                                                                   
the quark distributions are not modified in the low $Q^2$ region,\\             
{\tt IVAR=1,2}:                                                                 
they are made vanishing at low $Q^2$ through two                                
different modifications. They, as well as related physics,                      
are described in \cite{beseps}.\\                                               
{\tt ITERAD} \\                                                                 
{\tt ITERAD=0}:                                                                 
the quark distributions and $\alpha_{QED}$ (if running, see                     
below the description of flag {\tt IVPOL}) in leptonic bremsstrahlung           
are artificially chosen                                                         
to depend on $Q^2$ as defined from lepton momenta.                              
This option was implemented to have a possibility to compare results            
calculated by both TERAD and DISEP branches. \\                                 
{\tt ITERAD=1}:                                                                 
they depend on $Q_h^2$ as defined from the hadronic momenta                     
($Q_h^2$ is integration variable).                                              
This option must be used for realistic calculations. \\                         
{\tt IHCUT} \\                                                                  
{\tt IHCUT=0}: no cut on the hadronic final state, \\                           
{\tt IHCUT=1}: rejects hadronic final states with                               
$Q_h^2 \leq$ {\tt TCUT} $({\rm GeV}^2)$                                         
and invariant hadronic mass (photon not included)                               
$W_h^2 \leq$ {\tt AMF2CT} $({\rm GeV}^2)$. In this case the two desired         
cut values {\tt TCUT} and {\tt AMF2CT} are set. \\                              
{\tt IGCUT} \\                                                                  
{\tt IGCUT=0}: no cut on photon variables, \\                                   
{\tt IGCUT=1}: rejects all final states with a photon energy                    
$E_\gamma \geq E_{cut} ({\rm GeV})$                                             
and with photon angle in the interval $(\theta_{cut2},\theta_{cut1})$,          
angles in radians $(0, \pi )$ with respect to the electron beam. \\             
These two cuts ({\tt IHCUT} and {\tt IGCUT})                                    
are implemented only in {\tt DXYLEP}. \\                                        
{\tt IWEAK} \\                                                                  
This flag governs how EWRC are applied in the TERAD branch. \\                  
{\tt IWEAK=0}: they are applied only via $ \sin^2 \theta_W$, \\                 
{\tt IWEAK=1}: they are applied via the full set of the EW form factors;        
a QPM treatment is used for the structure functions $G_i$ and $H_i$             
in this case. \\                                                                
{\tt IZERO} \\                                                                  
a flag to swich on/off a time comsuming {\tt CALL} of the                       
TERAD branch in {\tt DXYLEP}, \\                                                
{\tt IZERO=0}: it is switched off; then the numbers of the TERAD                
branch are set to zero, \\                                                      
{\tt IZERO=1}: it is switched on.\\                                             
                                                                                
\subsection{DISEP - flags}                                                      
                                                                                
{\tt INORM}  \\                                                                 
{\tt INORM=0}: the nine                                                         
different QED corrections for the $\rm NC$ reaction                             
\cite{disepnc}                                                                  
are all normalized by the $ \gamma$-exchange Born cross section,\\              
{\tt INORM=1}: they are normalized by the complete Born cross section,          
$ \gamma+Z$-exchange,  \\                                                       
{\tt INORM=2}:                                                                  
each QED correction is normalized by its corresponding Born                     
contribution, three with $ \gamma$- exchange,                                   
three with $ \gamma Z$-interference                                             
and three with $Z$-exchange (three means:                                       
lepton, lepton-quark interference, and quark bremsstrahlung).                   
\\                                                                              
{\tt IEWRC} \\                                                                  
{\tt IEWRC=0}: no weak loop corrections are taken into account,\\               
{\tt IEWRC=1}: they are taken into account. \\                                  
{\tt IBOXF} \\                                                                  
{\tt IBOXF=1}:                                                                  
the contributions from $ZZ$- and $WW$-boxes are included,\\                     
{\tt IBOXF=0}: they are not included. \\                                        
(The {\tt IBOXF} flag acts only if {\tt IEWRC} $ \neq 0$.) \\                   
{\tt ICONV} \\                                                                  
{\tt ICONV=1}: weak form factors are running, \\                                
{\tt ICONV=0}:                                                                  
they do not run. The recommended value of this flag is 0. \\                    
(The {\tt ICONV} flag is active only for {\tt IEWRC=1}). \\                     
{\tt IQ20} \\                                                                   
{\tt IQ20=0}: the light quark masses will be replaced by a $Q^2_0$,             
which is recommended to be chosen equal to the                                  
$Q^2_0$ which enters the QCD evolution equations,                               
  \\                                                                            
{\tt IQ20=1}: then the light quark masses are used.\\                           
The first choice is recommended, the second is supported for the sake           
of comparison. \\                                                               
                                                                                
\subsection{Flags common both to TERAD and DISEP branches}                      
                                                                                
{\tt IEXP} \\                                                                   
{\tt IEXP=1}:                                                                   
the soft photon part of leptonic QED corrections is exponentiated,   \\         
{\tt IEXP=0}: it is not exponentiated.  \\                                      
{\tt IVPOL} \\                                                                  
{\tt IVPOL=0}: no running of $ \alpha_{QED}$,                                   
$ \alpha_{QED} = \alpha $,\\                                                    
{\tt IVPOL=1}: running $ \alpha_{QED}$.   \\                                    
{\tt IMOMS} \\                                                                  
This flag defines the calculational scheme used, i.e.                           
which EW parameters are considered as input quantities and which one            
is calculated (iterated). \\                                                    
{\tt IMOMS=1}: input - $ \alpha, G_\mu, M_Z, m_t, M_H $; $M_W$                  
is iterated,  \\                                                                
{\tt IMOMS=2}: input - $ \alpha, G_\mu, M_W, m_t, M_H $; $M_Z$                  
is iterated,  \\                                                                
{\tt IMOMS=3}: input - $ \alpha, M_W, M_Z, m_t, M_H $; $G_\mu$                  
is calculated. \\                                                               
It is recommended to use the {\tt IMOMS=1} option. Other options are            
introduced for some tests.  \\                                                  
                                                                                
There are also some additional parameters which could                           
be set by the user.                                                             
For the TERAD branch one can set parameters                                     
{\tt GG,GZ,ZZ (=0,1)} which switch on/off the contributions due to              
$\gamma$-exchange, $ \gamma Z$-interference, $Z$-exchange, respectively,        
in cross sections in any variable in the TERAD branch.                          
(Together with {\tt INORM=2} one must fix {\tt GG=GZ=ZZ=1}.)                    
                                                                                
There are two parameters {\tt W2TR} and {\tt TTR}                               
          used to subdivide the 2-dimensional                                   
integration region in all {\tt DXY...}'s into up to three parts.                
For {\tt W2TR=TTR=0}, there is only one region.                                 
Choosing the parameter {\tt TTR} (in ${\rm GeV}^2$) $ > 0$, one                 
splits the region into two parts, one of them with small $Q_h^2$                
where the structure functions should have a special behaviour.                  
In addition, this region of small $Q_h^2$ can be split into                     
two subregions                                                                  
by a choice of the parameter {\tt W2TR} (in ${\rm GeV}^2$)                      
$ > M_{proton}^2$.                                                              
The latter was used for a proper treatment of the resonance region.             
Such subdivision of the kinematical region is used for {\tt MODEL=0}.           
For more details about the low $x$ region see \cite{dimajacek}.                 
                                                                                
Two more parameters are: {\tt ALAM}                                             
 - the lepton beam polarization (in the interval $(-1,+1)$), and                
{\tt LEPCH} -                                                                   
the lepton beam charge, $-1$ for electron, $+1$ for positron beams              
respectively.                                                                   
                                                                                
In {\tt SUBROUTINE SET} one sets also some constants like $ \pi$,               
proton and electron masses,                                                     
electromagnetic coupling constant $ \alpha$, Fermi coupling $G_\mu$,            
and some combinations of them. \\                                               
                                                                                
Then goes a part which initializes the calculation of                           
electroweak parameters. It needs an additional setting                          
of flags (they are stored in the {\tt NPAR}-array)                              
and constants for initialization of the EW library {\bf DIZET}.                 
Since this library was originally                                               
designed for LEP-I physics some flags and                                       
parameters have not too much meaning at HERA, therefore for some of             
them we simply report the recommended value avoiding further                    
explanation. \\                                                                 
{\tt IHVP}  \\                                                                  
says how to calculate the hadronic vacuum polarization (HVP)                    
contribution to the running $ \alpha_{QED}$: \\                                 
{\tt IHVP=1}: HVP of Jegerlehner, \cite{jegerlehner},\\                         
{\tt IHVP=2}: HVP is made of some effective quark masses, \\                    
{\tt IHVP=3}: HVP of Burkhardt et al., \cite{burkhardt}, \\                     
{\tt IHVP=3}: recommended value.\\                                              
{\tt IAMT4} \\                                                                  
says how to apply some leading higher order corrections,                        
connected with the top-quark, to $Z,W$ widths and EW form factors: \\           
{\tt IAMT4=0}: they are neglected, \\                                           
{\tt IAMT4=1,2,3}: different variants of their treatment, \\                    
{\tt IAMT4=3}: recommended value. \\                                            
{\tt IQCD} \\                                                                   
says how to apply QCD ${ \cal O} ( \alpha \alpha_s)$ corrections:   \\          
{\tt IQCD=0}: they are neglected,    \\                                         
{\tt IQCD=3}: they are included approximately                                   
(applicable only at LEP-I), \\                                                  
{\tt IQCD=4}: they are calculated completely,    \\                             
{\tt IQCD=4}: recommended value                                                 
(with a subsequent redefinition, see below). \\                                 
{\tt IMASS} \\                                                                  
{\tt IMASS=0}: recommended value. \\                                            
{\tt IALST} \\                                                                  
{\tt IALST=1}: recommended value. \\                                            
{\tt IQCD3} \\                                                                  
{\tt IQCD3=1}: recommended value. \\                                            
                                                                                
The $Z$-boson mass {\tt AMZ}, top-quark mass {\tt AMT},                         
Higgs boson mass {\tt AMH} and                                                  
{\tt VARQCD} $= \alpha_s(M_Z^2)$ (=0.12 recommended value)                      
should be also set for the subsequent call of the                               
EW library, {\tt CALL DIZET}.                                                   
For {\tt IMOMS=1, DIZET} calculates the $W$-boson mass {\tt AMW}, but           
having in mind {\tt IMOMS} $ \neq 1$ options, the $W$-mass                      
can also be set. In the {\tt ZPAR}-array,                                       
{\tt DIZET} returns many useful static EW                                       
quantities. We mention only two: $ \Delta r$ is stored in                       
{\tt ZPAR(1)} and $ \sin^2 \theta_W$ in {\tt ZPAR(3)}.                          
                                                                                
After the {\tt DIZET} call one could                                            
re-initialize, if needed, the EW parameters. To do that it is sufficient        
to redefine {\tt AMZ} and {\tt AMW}. There is then one more important           
redefinition. As the {\tt IQCD=4} option is very time consuming, we             
recommend to retain after the call of {\tt DIZET} two statements:\\             
{\tt IQCD = 0}          \\                                                      
{\tt NPAR(3) = IQCD}       \\                                                   
which will result in applying ${ \cal O} ( \alpha \alpha_s)$ corrections        
only in $ \Delta r$ but not in the EW form factors. For applications            
at HERA this is a very good approximation. The subsequent                       
{\tt CALL CONSTQ} initializes all fermion (lepton and quark) masses             
and charges.                                                                    
                                                                                
Then one sets some other constants like                                         
{\tt QE} $=2I_3^e$, twice the weak isospin of the electron,                     
{\tt GV} and {\tt GA}, vector and axial vector electron couplings               
to the $Z$-boson, and the pion threshold                                        
{\tt W2PIT} $=(M_{proton}+M_{pion})^2$.                                         
                                                                                
{\tt SUBROUTINE SET}                                                            
ends with the calculation of the $s$-invariant at HERA and                      
the creation of a lattice in $x$ and $y$                                        
over the kinematic domain in which DIS cross sections will be calculated.       
The version described creates the 6($x$)*16($y$)=96 points                      
defined by \\                                                                   
\begin{center}                                                                  
$x=.0001,.001,.01,.1,.5, .9$ \\                                                 
and \\                                                                          
$y=.01,.02,.05,.1,.15,.2,.3,.4,.5,.6,.7,.8,.9,.95,.98,.99$.\\                   
\end{center}                                                                    
\vspace{.5cm}                                                                   
                                                                                
\section{Further developments of the code}                                      
                                                                                
The code {\bf TERAD91} was frozen in October '91 and used to produce            
numbers for various Figures and Tables of the Workshop Proceedings.             
Last bug fixes were made on February 10th of 1992, for this reason              
it is recommended to use a version saved after that date.                       
Nowadays only one new calculation is in progress: the DZRCG creates             
a DISEP branch for the $\rm NC$ reaction in mixed and hadronic variables.       
Later it is planned to add a                                                    
DISEP branch for the $\rm CC$ reaction in                                       
hadronic variables. Further                                                     
efforts in the future are not excluded if some new results will be              
requested by HERA experiments. \\                                               
\vspace{.5cm}                                                                   
                                                                                
The {\bf TERAD91} package consists of about 11150 lines of coding               
(including about 4000 lines of weak library {\bf DIZET}).                       
                                                                                
                                                                                
                                                           

\begin{thebibliography}{99}                                                     
                                                                                
\bibitem{contvol1}                                                              
H. Spiesberger, A. Akhundov et al., these Proceedings, Vol. I.                  
\bibitem{teradtex}                                                              
A. Akhundov, D. Bardin, L. Kalinovskaya, T. Riemann,                            
{\it "Model-independent QED corrections to the process $ep \rightarrow          
eX$ at HERA energies", a complete description of the TERAD approach,            
in preparation.}                                                                
\bibitem{disepnc}                                                               
D. Bardin, C. Burdik, P. Christova, T. Riemann,                                 
Z. Physik {\bf C42} (1989) 679.                                                 
\bibitem{disepcc}                                                               
D. Bardin, C. Burdik, P. Christova, T. Riemann,                                 
Z. Physik  {\bf C44} (1989) 149.                                                
\bibitem{dizet}                                                                 
A. Akhundov, D. Bardin, M. Bilenky, P. Christova,                               
S. Riemann, T. Riemann, M. Sachwitz, H. Vogt,                                   
version 4.04 (21 Aug 1991);                                                     
version 2 was described in:                                                     
    D. Bardin et al.,                                                           
    Comput. Phys. Commun. {\bf 59} (1990) 303;                                  
a description of version 4.04 is in preparation.                                
\bibitem{oldpreps}                                                              
     A. Akhundov, D. Bardin, JINR Dubna prepr. P2-9587 (1976)                   
     (in Russian); \\                                                           
     A. Akhundov, D. Bardin, N. Shumeiko,                                       
     JINR Dubna prepr. E2-10147 (1976);      \\                                 
     A. Akhundov, D. Bardin, N. Shumeiko,                                       
     JINR Dubna prepr. E2-10205 (1976).                                         
\bibitem{oldpub}                                                                
A. Akhundov, D. Bardin, N. Shumeiko,                                            
Sov. J. Nucl. Phys. {\bf 26} (1977) 660.                                        
\bibitem{terad86}                                                               
A. Akhundov, D. Bardin, W. Lohmann, JINR Dubna prepr. E2-86-104 (1986).         
\bibitem{bcdms}                                                                 
A. Akhundov et al., CERN-EP/89-06; Phys. Lett. {\bf B223} (1989) 485.           
\bibitem{cdhsw}                                                                 
H. Abramowicz et al., Z. Physik, {\bf C17} (1983) 283.                          
\bibitem{charmI}                                                                
F. Bergsma et al., CERN-EP/85-113, Contribution to the Int. Symp.               
on Lepton and Photon Interactions at High Energies, Kyoto, 19-24 August         
1985.                                                                           
\bibitem{dukeowens}                                                             
D. W. Duke and J. F. Owens, Phys. Rev. {\bf D30} (1984) 49.                     
\bibitem{brasse}                                                                
F. W. Brasse et al., Nucl. Phys. {\bf B110} (1976) 413.                         
\bibitem{stein}                                                                 
S. Stein et al., Phys. Rev. {\bf D12} (1975) 1884.                              
\bibitem{dimajacek}                                                             
D. Bardin and J. Ciborowski, these Proceedings, Vol. I.                         
\bibitem{beseps}                                                                
A. Akhundov, D. Bardin, L. Kalinovskaya, Z. Physik {\bf C51} (1991) 557.        
\bibitem{jegerlehner}                                                           
F. Jegerlehner, PSI-PR-91-16 (1991),                                            
in: Progress in Particle and Nuclear                                            
Physics, ed. A. Fassler, Pergamon press, Oxford, UK, {\bf vol.27}, p.1.         
\bibitem{burkhardt}                                                             
H. Burkhardt, F. Jegerlehner, G. Penso, C. Verzegnassi,                         
Z. Physik {\bf C43} (1989) 497.                                                 
\end{thebibliography}
\end{document}